\documentclass{webofc}
\usepackage[varg]{txfonts}   
\usepackage{listings}
\usepackage{hyperref}
\usepackage[inline]{trackchanges} 
\addeditor{LPG} 
\addeditor{LE} 
\addeditor{LR} 
\addeditor{BH} 
\addeditor{MAU} 
\addeditor{NM} 

\begin{document}
\title{Pulsating star research and the Gaia revolution.}
%
%

\author{\firstname{Laurent}   \lastname{Eyer}\inst{1}\fnsep\thanks{\href{mailto:Laurent.Eyer@unige.ch}{\tt Laurent.Eyer@unige.ch}} \and
        \firstname{Gisella}   \lastname{Clementini}\inst{2} \and
        \firstname{Leanne P.} \lastname{Guy}\inst{3}        \and
        \firstname{Lorenzo}   \lastname{Rimoldini}\inst{3}  \and
        \firstname{Florian}   \lastname{Glass}\inst{1}      \and
        \firstname{Marc}      \lastname{Audard}\inst{1,3}   \and
        \firstname{Berry}      \lastname{Holl}\inst{1,3}    \and
        \firstname{Jonathan}   \lastname{Charnas}\inst{3}   \and                 
        \firstname{Jan}        \lastname{Cuypers\footnote{Deceased on February 28, 2017}}\inst{4}     \and
        \firstname{Joris}      \lastname{De Ridder}\inst{5}               \and
        \firstname{Dafydd W.}  \lastname{Evans}\inst{6}                   \and
        \firstname{Gregory}    \lastname{Jevardat de Fombelle}\inst{3,7}  \and
        \firstname{Alessandro} \lastname{Lanzafame}\inst{8,9}             \and
        \firstname{Isabelle}   \lastname{Lecoeur-Taibi}\inst{3}           \and
        \firstname{Nami}       \lastname{Mowlavi}\inst{1,3}               \and
        \firstname{Krzysztof}  \lastname{Nienartowicz}\inst{3,7}          \and
        \firstname{Marco}      \lastname{Riello}\inst{6}                  \and
	\firstname{Vincenzo}   \lastname{Ripepi}\inst{10}                 \and
        \firstname{Luis}       \lastname{Sarro}\inst{11}                  \and
	\firstname{Maria}      \lastname{S\"{u}veges}\inst{3,12}         
}

\institute{
	Department of Astronomy, University of Geneva, Ch. des Maillettes 51, 1290 Versoix, Switzerland
\and
	INAF - Osservatorio Astronomico di Bologna, Via Gobetti 93/3, 40129 Bologna, Italy
\and
        Department of Astronomy, University of Geneva, Ch. d’Ecogia 16, 1290 Versoix, Switzerland  
\and
	Royal Observatory of Belgium, Ringlaan 3, 1180 Brussels, Belgium
\and
	Institute of Astronomy, KU Leuven, Celestijnenlaan 200D, 3001 Leuven, Belgium
\and
        Institute of Astronomy, University of Cambridge, Madingley Road, Cambridge CB3 0HA, UK
\and
	SixSq, Rue du Bois-du-Lan 8, 1217 Geneva, Switzerland
\and
	Universita di Catania, Dipartimento di Fisica e Astronomia, Sezione Astrofisica, Via S. Sofia 78, 95123 Catania, Italy
\and
	INAF-Osservatorio Astrofisico di Catania, Via S. Sofia 78, 95123 Catania, Italy
\and
	INAF-Osservatorio Astronomico di Capodimonte, Via Moiariello 16, 80131 Napoli, Italy
\and
	Dpto. Inteligencia Artificial, UNED, c/ Juan del Rosal 16, 28040 Madrid, Spain
\and
	Max Planck Institute for Astronomy, Koenigstuhl 17, 69117 Heidelberg, Germany
          }

\abstract{%
In this article we present an overview of the ESA Gaia mission and of the unprecedented impact that Gaia will have on the field of variable star research. 
We summarise the contents and impact of the first  Gaia data release on the description of variability phenomena, with particular emphasis on pulsating star research. 
The Tycho-Gaia astrometric solution, although limited to 2.1 million stars, has been used in many studies related to pulsating stars. Furthermore a set of 3,194 Cepheids and RR Lyrae stars with their times series have been released.
Finally we present the plans for the ongoing study of variable phenomena with Gaia and highlight some of the possible impacts of the second data release on variable, and specifically, pulsating stars.
}
\maketitle
%

\section{Introduction}\label{sect:introduction}

The variable phenomena of celestial objects present very diverse behaviour.
An attempt to visualise this zoo of variability types was presented in \protect \cite{EyerMowlavi2008} and shows that pulsating stars form an abundant group.

Gaia, as a global and comprehensive survey, will have an unprecedented impact on the field of pulsating star research and more generally on variable star research.
Gaia will be able to detect many/most of the variability types presented in the variability tree of \cite{EyerMowlavi2008}.
Furthermore, this mission will help to characterise most if not all stars that are members of a variability type presented in this tree.

More concretely, we can enumerate some examples of the foreseen impact of Gaia:
\begin{enumerate}
\item The discovery of an enormous number of new variable stars,
\item The definition of the precise location of instability strips in the Colour Magnitude Diagram (CMD),
\item The calibration of standard candles,
\item  The comparison of results with those obtained from asteroseismology,
\item  The description of structures of the Milky Way and other galaxies with variable stars used as tracers of certain galactic populations.
\end{enumerate}

Gaia's three instruments observe objects near-simultaneously to collect astrometric, photometric and spectroscopic data, covering the entire sky.
This multi-messenger aspect of Gaia is a strong asset for the study of variability as compared with single band surveys.
We can remark that currently the number of stars known with multiple pieces of information (good astrometric precision, photometric and radial velocity time series) is very low.
There are only 30 pulsating stars known with a relative parallax precision of better than 10\% (from Hipparcos~\cite{PerrymanEtAl1997}), Hp-band photometric and Coravel~\cite{BaranneEtAl1979} radial velocity time-series.
It is clear that Gaia will increase these figures by several orders of magnitude.

Gaia will be, for example, particularly interesting when we find pulsating a star in a binary system that is a member of a cluster.
Gaia may be able to characterise the "different" objects, i.e. the cluster, the binary system, the star and the pulsation.
The comparison of these different and independent pieces of information can provide stringent tests for our understanding of pulsation and stellar evolution.

Obviously Gaia won't solve all problems; there are limitations on the precision of measurements and also in the density of the time sampling.
The combination of Gaia with other surveys, with other types of measurements (e.g. high cadence photometry, high resolution spectroscopy, interferometry) and other methods of determining distance will be another way to significantly increase the scientific impact of Gaia on pulsating star research.

Within the Gaia consortium, a dedicated group is charged with the variability processing and analysis, see \cite{EyerEtAl2017}. 
This analysis is mostly based on the work of the group responsible for the photometric reduction \cite{vanLeeuwenEtAl2016}.
The results of this variability processing was and will be made public across official ESA data releases. 
In the two following sections we explain the content of the first and second data releases.

\section{Gaia Data Release 1}\label{sect:GDR1}

\subsection{The contents of Gaia Data Release 1}

The contents of the first Gaia data release were based on the accumulation of 14 months of Gaia observations, collected between 25 July 2014 and 16 September 2015.
During these 14 months, two different scanning laws were employed. 
For an initial period of 28~days, the Ecliptic Poles Scanning Law (EPSL) (which repeatedly covered the ecliptic poles on every spin) was used, producing high-cadence measurements in these regions.
During the subsequent 13 months the Nominal Scanning Law (NSL) was used.

This 14-month data set comprises more than half a trillion observations that were processed and used to derive the following contents of this first Gaia public release:
\begin{itemize}
\item Positions and mean G-band photometry for 1.1~billion stars,
\item Parallaxes and proper motions for 2.1~million stars,
\item Light curves of 3,194~Cepheids and RR~Lyrae stars,
\item Positions and magnitudes of 2,191~quasars.
\end{itemize}

The Gaia consortium produced comprehensive documentation to accompany the first data release, which can be found on the ESA website (\url{https://gaia.esac.esa.int/documentation/GDR1/}).
A series of articles were published in the journal Astronomy \& Astrophysics.
The mission is described in \cite{PrustiEtAl2016} and the results of the first release are summarised in \cite{BrownEtAl2016}.
For the topics related to variable stars and stellar pulsation, the interested reader can find specific information in the following three articles:
\begin{enumerate}
\item On the general approach of the variability analysis developed by the Gaia consortium and its application to G-band time series, \cite{EyerEtAl2017}.
      Note that this analysis, aimed at extracting a list of Cepheid and RR~Lyrae candidates, was restricted to objects within 38~deg of the South Ecliptic Pole.
\item On the specific processing applied to the Cepheid and RR Lyrae candidates, to derive parameters from a Fourier decomposition of the light curve \cite{ClementiniEtAl2016}.
\item On the period luminosity relation of Cepheids and RR Lyrae stars \cite{ClementiniEtAl2017} derived from the Tycho-Gaia Astrometric Solution (TGAS) \cite{LindegrenEtAl2016}.
\end{enumerate}

For a mission of this breadth and complexity, to have produced a first data release of this quality only 2 years following the beginning of spacecraft operations is a real achievement.
Obviously, there are some limitations, for which the reader should refer to the official website for the first Gaia data release (\url{http://www.cosmos.esa.int/web/gaia/dr1}). 
We list here a few of the limitations which could be pertinent to pulsating star research:
\begin{itemize}
 \item Some filtering has been applied, e.g. cuts on the number of observations, cuts on colour, cuts on objects with "too large" uncertainties, cuts on sources lacking astrometry and/or photometry, cuts on sources with an upper limit on errors in parallax, position and photometry, etc.
 \item There are no high-proper-motion stars (i.e. with $\mu > 3.5$ arcsec/year),
 \item All sources are treated as single stars (so the parallax of some astrometric binary stars may obviously be erroneous), 
 \item Various un-modelled effects are present in the data (chromaticity, micro-meteoroid hits, micro-clanks, \ldots),
 \item Variations of the basic angle were derived from the on-board metrology,
 \item There are some spurious sources.
\end{itemize}
One consequence of these limitations is that your favorite source may be missing from the data release.
It is important to bear in mind that the processing of the Gaia data by the DPAC consortium is iterative;
with each subsequent data release the quality of the data improves as additional effects are taken into account or effects are better modelled.

\subsection{The processing and analysis of Cepheids and RR~Lyrae stars}

The classification selection of RR~Lyrae stars and Cepheids was done using supervised classification algorithms \cite{EyerEtAl2017}.
These classifiers were expressly tuned to be permissive. 
We remark that for the South Ecliptic Pole region, the completeness is very high.
A comparison with the OGLE survey on RR~Lyrae stars and Cepheids showed that all stars observed by the OGLE survey have also been observed by Gaia.
Following the application of various selection and quality criteria, including 
{(\it i)} a cut on the number of measurements to be greater than 20,
{(\it ii)} a requirement that the star be processed in two different reduction cycles, 
{(\it iii)} the removal of stars for which an unambiguous definition of the period and a robust classification as Cepheid or RR~Lyrae was not possible, and
{(\it iv)} the removal of multimode stars,
we arrive at a completeness of 58\% for RR~Lyrae stars and 67\% for Cepheids.
These numbers are similar to those found by \cite{UdalskiEtAl2016}.

The inclusion of Cepheids and RR~Lyrae stars in the first Gaia data release should be considered a bonus as it was originally not scheduled until much later in the DPAC processing cycle.
Our primary goal was to showcase a selection of stars somewhat similar (in terms of the number of measurements) to what we expect in the final data release.

\begin{figure*}
\centering
\includegraphics[width=10cm,clip]{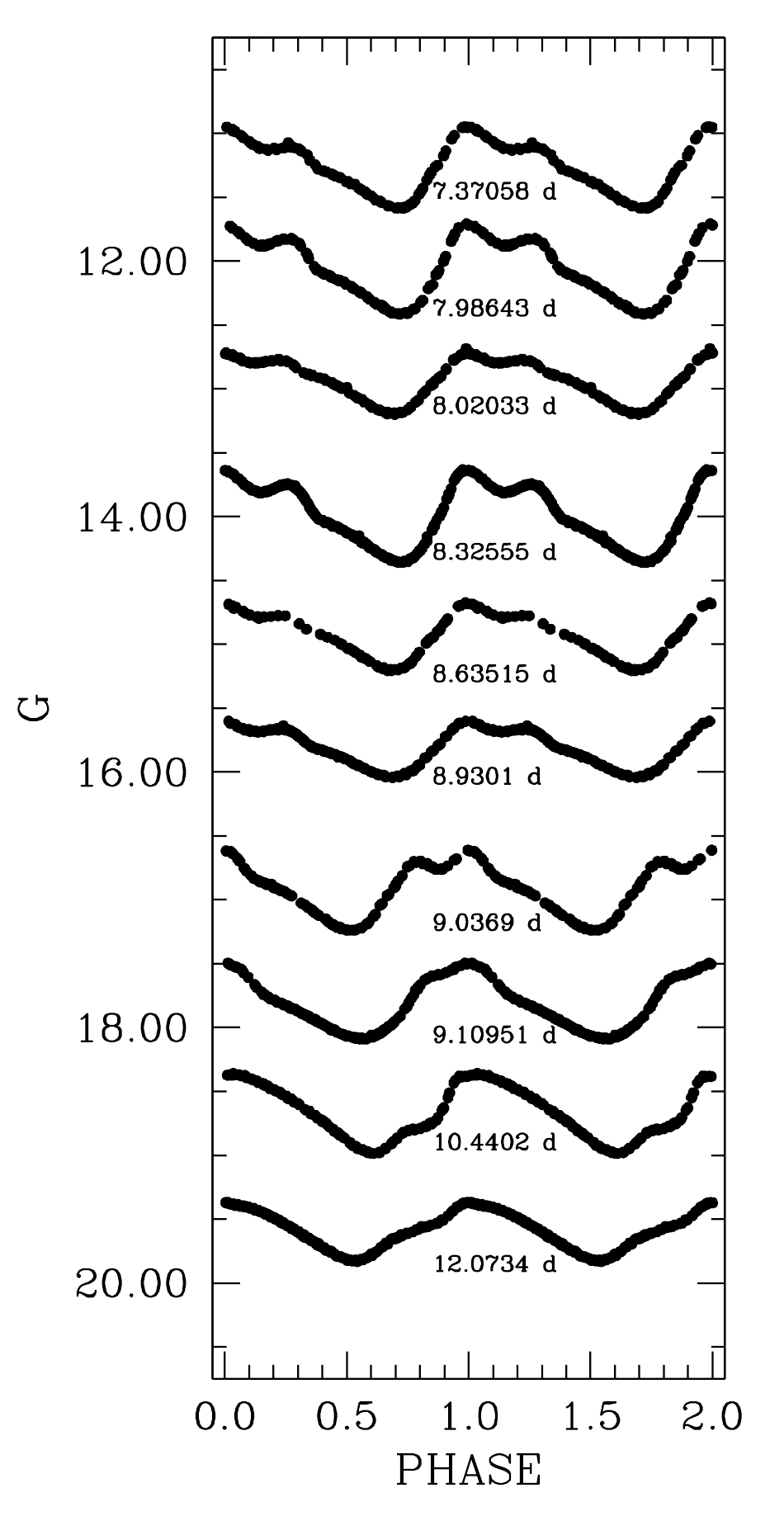}
\caption{The Hertzsprung progression of classical Cepheids with Gaia data.}
\label{fig:CepheidHP}       
\end{figure*}

The quality of the photometry allows a more detailed description of the light curve.
An approach to describe light curve morphology was done through a Fourier series and was published in the first data release, see \cite{ClementiniEtAl2016}.
We can remark on the quality of the photometry when we present for example the Hertzsprung progression \cite{Hertzsprung1926}.
This phenomenon, the period-dependence of the location of a bump in certain Cepheid light curves, is shown in Figure~\ref{fig:CepheidHP}.  

\subsection{The structure of Magellanic Clouds thanks to pulsating stars} \label{subsect:RRLyraestars}
It is well known that some pulsating stars can be used as standard candles and tracers of certain galactic populations.
Among them are the Cepheids, RR~Lyrae stars and Mira stars.

From the first Gaia data release, it could be remarked that the new Large Magellanic Cloud RR~Lyrae stars discovered in the South Ecliptic Pole region seem to trace a more extended halo than previously thought \cite{ClementiniEtAl2016}.

The OGLE team produced 2 articles about the structure of the Magellanic Clouds \cite{Jacyszyn-DobrzenieckaEtAl2016a, Jacyszyn-DobrzenieckaEtAl2016b}.
The second article was contemporaneous to a study made by \cite{BelokurovEtAl2016} which used the mean photometry of the first Gaia data release to extract RR~Lyrae star candidates in the neighbourhood of the Magellanic Clouds.
\cite{BelokurovEtAl2016} made comparisons between the different studies and remark that a bridge of such stars exists between the Large and Small Magellanic Clouds and conclude that it is a stellar tidal tail. 
Using a similar method with Gaia data but using Mira stars instead of RR~Lyrae stars, \cite{DeasonEtAl2016} are unveiling the outskirts of the Large Magellanic Cloud. 
Although based on a small fraction of the Gaia data, such studies demonstrate the power of using certain groups of pulsating stars to describe the structures of two of our neighbouring galaxies.

\subsection{Tests of distances and parallaxes}

There has been several comparisons between the Gaia parallaxes and distances obtained from pulsating/variable stars:

For example Cepheids were used by \cite{LindegrenEtAl2016}, to select far away objects to test the zero point of the Gaia parallaxes and the north-south asymmetry.
Both are found to be unsignificant. A series of comparisons was performed by \cite{ClementiniEtAl2017} with Hipparcos and HST parallaxes.
Furthermore, Clementini et al. (this volume and \cite{ClementiniEtAl2017}) derived new Period-Luminosity relations.
Alternatively \cite{CasertanoEtAl2017} used 212 Galactic Cepheids to test the Gaia parallaxes. They conclude that the Gaia parallax uncertainties seem to be overstimated.

A series of articles used asteroseismic results to derive distances that can be compared to Gaia parallaxes. 
Here comes some statistical difficulties to compare on one side parallaxes and distances, especially when $\mid \frac{\sigma_\pi}{\pi} \mid$ becomes large.
It should be noted that such comparison, when concordant, validates both the asteroseismic distance and parallaxe determinations.
Such studies can be found in the articles \cite{DeRidderEtAl2016}, \cite{SilvaAguirreEtAl2017}, \cite{DaviesEtAl2017}.
When statistically significantly different, a detailed and thorough study should be done to understand the discrepancy.
Currently the second data release will be much more precise that the TGAS parallaxes, so that we are eagerly waiting it to understand some of these discrepancies.

Variable stars other than pulsating stars can be used as tests of the Gaia parallaxes. For example \cite{StassunTorres2016} used eclipsing binaries.

\subsection{Position of variable stars in the Colour Magnitude Diagram}

We used the TGAS parallaxes \cite{LindegrenEtAl2016} as well as the $B_T-V_T$ Tycho colours \cite{HoegEtAl2000} to construct a CMD.
We selected stars that have $\mid \frac{\sigma_\pi}{\pi} \mid < 20\%$ and with a standard deviation on the colour of less than 0.1 mag.
We took care not to exclude some large amplitude variables that could have large colour variations and would therefore inflate the value of the standard deviation.
We used the AAVSO catalogue of variable stars \cite{Watson2006} to crossmatch with the Tycho identifiers.

\begin{figure*}
\centering
\includegraphics[width=13cm,clip]{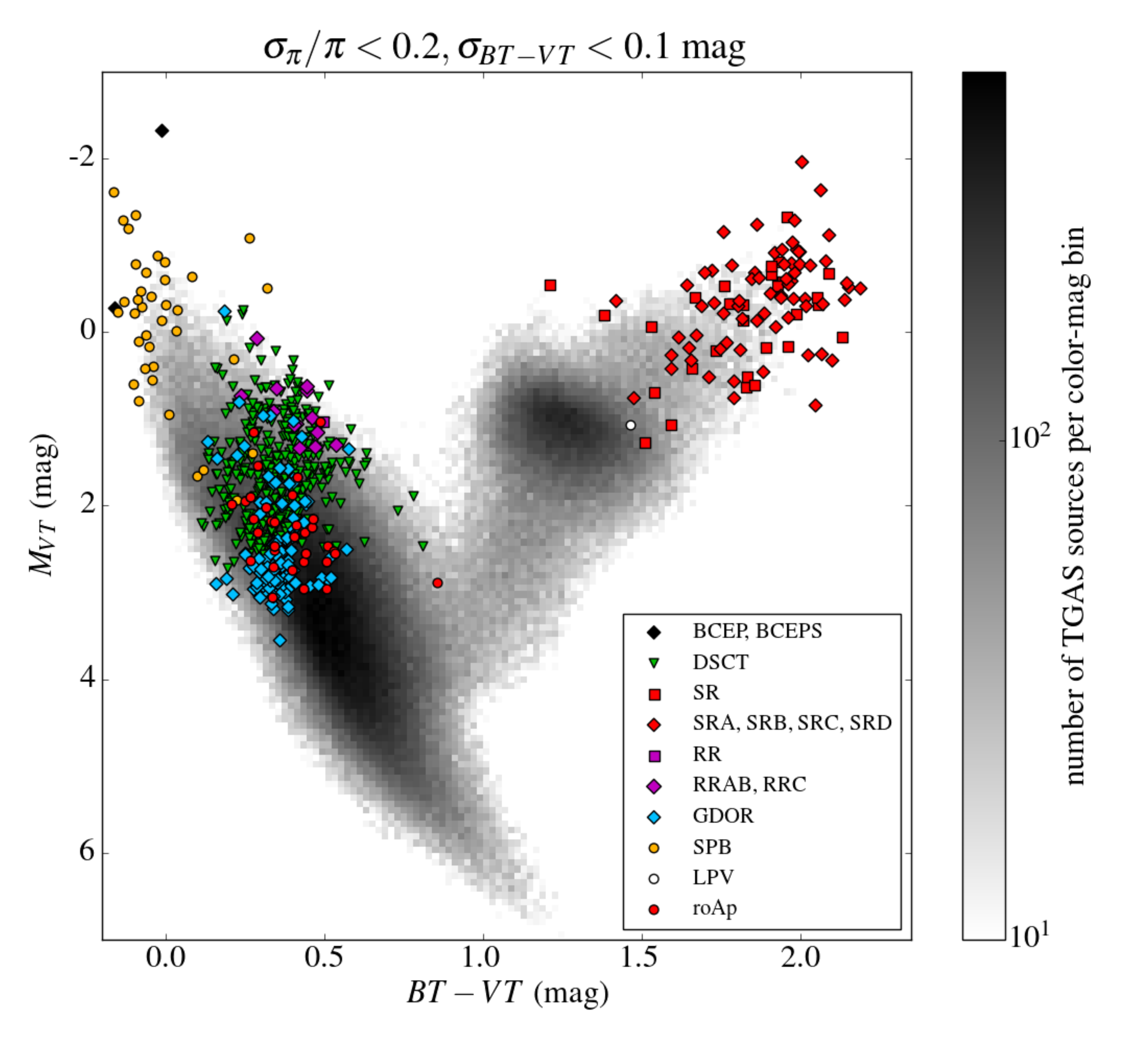}
\caption{Colour Magnitude diagram showing the location of several pulsating stars.}
\label{fig:CMDVarTypes}       
\end{figure*}

The quality of the AAVSO catalogue, being a compilation of many heterogenous sources, can vary considerably. 
Many crossmatch problems and incorrect class assignments resulting from single-band photometric surveys, especially near the Galactic plane, were identified and removed from the used data. 
We present the resulting CMD in Figure~\ref{fig:CMDVarTypes}. 
Such a diagram, when fully cleaned using data from future Gaia data releases, will be most useful in determining regions of variability of certain variability types and can be used to compare theoretical predictions of the instability strips.  Working on this subject, with the encountered problems, we can state that we eagerly await future Gaia data releases, which will reduce many of the problems we have had to face here.
Indeed Gaia has the advantage of having one set of instruments observing the whole sky and therefore producing an homogenous and coherent data set.

\subsection{Calibration of luminosities of standard candles}

Gaia will play a crucial role in the calibration of different kinds of standard candles.

For the first Gaia data release a study was done by Clementini et al. (\cite{ClementiniEtAl2017}, see also this volume).
Given the precision of the TGAS astrometry, the best improvement was seen for RR~Lyrae star luminosity calibrations.
Clementini et al. also showed the differences between different statistical methods used to calibrate the RR~Lyae stars, Classical Cepheids and Type~II Cepheids.
The three different methods are: a direct least square applied to the log(period)-magnitude diagram, the ABL method \cite{ArenouLuri1999} and a Bayesian approach. 
The first method is known to introduce biases in the estimates of luminosity calibrations when $\mid \sigma_{\pi}/\pi \mid$ is large.
Biased estimators can occur because of many different reasons.
One way to get a biased estimator is to estimate a quantity from censored samples, here for example due to the removal of sources with negative measured parallaxes, for which no distance or absolute magnitude can be inferred. An other way to get a biased estimator is when uncertainties have distributions which differs (asymmetric and/or non gaussian distribution) from the desired ones.

Clementini et al. (this volume and \cite{ClementiniEtAl2017}) are showing that the direct "naive" least squares method gives a distance shorter than more elaborated methods.

We can predict that the second Gaia data release will have a profound impact on any of the pulsating stars which are standard candles (LPVs, OSARGs, Cepheids, RR~Lyrae stars, etc.).

\section{Future Data Releases}\label{sect:future}
In 2012, a release scenario was presented in a Gaia Consortium internal document by Wil~O'Mullane.
In that scenario five data releases were planned, and were presented for example in \cite{EyerEtAl2015}.
After the success of the first data release, and its associated knowledge, the long term planning of the future releases for the Gaia archive is currently under revision.
If the number of releases is under discussion, the goal for the content of the final release is still a target.

The second Gaia Data Release is planned for April 2018.
This new date, compared to the initial planned date, was driven by the desire of the Gaia consortium to have a self-consistent data release in which every delivered component makes use of the same reduction cycle data. 
The content of this second Data Release is described in \url{https://www.cosmos.esa.int/web/gaia/release} and reproduced below:

\begin{itemize}
 \item Five-parameter astrometric solutions for all sources with acceptable formal standard errors ($>10^9$ anticipated), and positions $(\alpha, \delta)$ for sources for which parallaxes and proper motions cannot be derived.
 \item G and integrated BP and RP photometric mean fluxes and magnitudes for all sources.
 \item Median radial velocities for sources brighter than $G_\mathrm{RVS}=12$ mag.
 \item For stars brighter than $G=17$ mag estimates of the effective temperature and, where possible, line-of-sight extinction will be provided, based on the above photometric data.
 \item Photometric epoch data for a sample of variable stars.
 \item Epoch astrometry for a pre-selected list of $>10,000$ asteroids.
\end{itemize}
The actual data release content may, however, deviate from the above target content.

\subsection{Variability outputs}
The Gaia variability team has set ambitious goals for the upcoming data releases. 
First, we aim to increase significantly the number of released objects compared to the first Data Release, up to two order of magnitudes and to increase the complexity of the data processing.
For the second data release, we aim to use the Gaia astrometric data and the BP and RP integrated photometry, and to release the G, BP, and RP photometric time series.
Several work packages, corresponding to distinct variability types, are activated for the second data release.

Apart from identifying different types of variability in Gaia data, one major goal is to identify at least RR~Lyrae stars over the entire celestial sphere.
Using data from the 14-month solution, we can achieve total coverage with less than 10 measurements.

\subsection{Variability types as tracers of galactic structure}
\begin{figure*}
\centering
\includegraphics[width=12cm,clip]{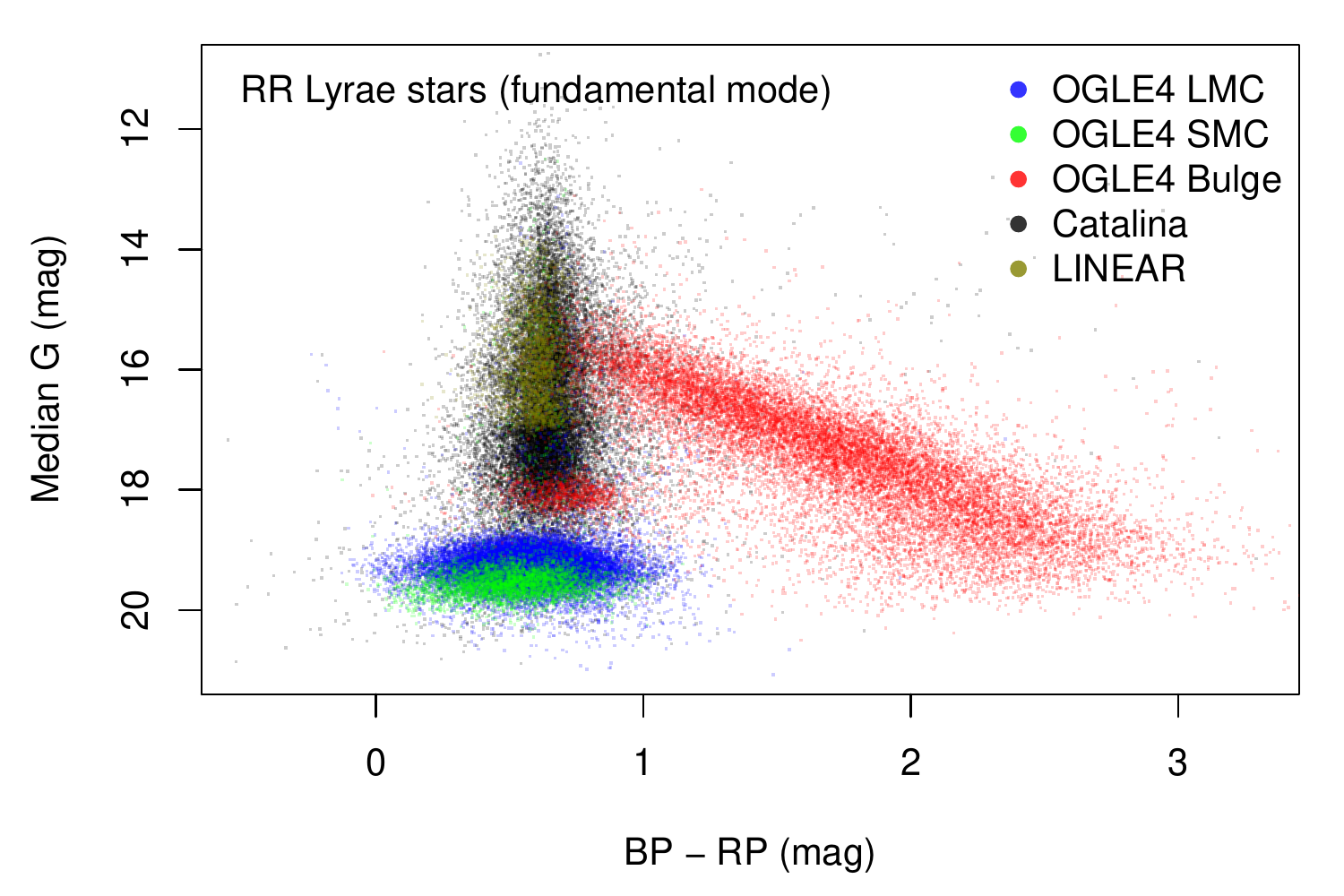}
\caption{Colour-(apparent) magnitude diagram, in Gaia pass-bands of known fundamental mode (ab-type)  RR Lyrae stars. Magnitudes and colour are not corrected for extinction (from the \href{https://www.cosmos.esa.int/web/gaia/iow_20170324}{Gaia Image of the Week of 24/03/2017}).}
\label{fig:ColorMagAForRRAB}       
\end{figure*}

Pulsating stars can be, and have been used to study galactic structures (see Sect.~\ref{subsect:RRLyraestars}). Figure~\ref{fig:ColorMagAForRRAB} shows a colour-magnitude diagram of 71,464 known fundamental mode ("ab" Bailey type) RR~Lyrae stars.

These known RR~Lyrae stars are coming from different surveys: OGLE \cite{SoszynskiEtAl2016, SoszynskiEtAl2014}, the Catalina survey \cite{DrakeEtAl2013a, DrakeEtAl2013b, DrakeEtAl2014, TorrealbaEtAl2015} and LINEAR survey \cite{PalaversaEtAl2013}.
We cross-matched these catalogues with the Gaia catalogue, allowing us to obtain an homogeneous set of magnitudes and colours from G, BP, and RP photometry.

This diagram of apparent magnitude versus colour reveals very interesting structures.
We discern in the diagram some over-densities, sub-structures. 
It would be most interesting to see if these over-densities are due to streams or are spatially clumped. 
We can see that
\begin{itemize}
 \item the black and brown points (Catalina and LINEAR surveys) producing a vertical structure that traces the Galactic halo.
 \item The diagonal structure in red is caused by interstellar extinction, which attenuates and reddens the light of (RR Lyrae) stars in the direction of the Bulge.
       These stars are also affected by metallicity, expected to increase the spread of apparent magnitudes by about 0.2-0.4 mag \cite{ClementiniEtAl2003}.
       The small clump in red (with G $\sim$ 18 mag) includes primarily RR Lyrae stars from the Sagittarius dwarf spheroidal galaxy at a heliocentric distance of about 25kpc.
 \item the blue and green points correspond to RR~Lyrae stars in the Large and Small Magellanic Clouds.
       Notice that the Large Magellanic Cloud is closer than the Small Magellanic Cloud, with a difference in distance modulus of about 0.5 mag, and both galaxies show a larger spread in colour indices (partially due to the photon noise).
\end{itemize}

\begin{figure*}
\centering
\includegraphics[width=14cm,clip]{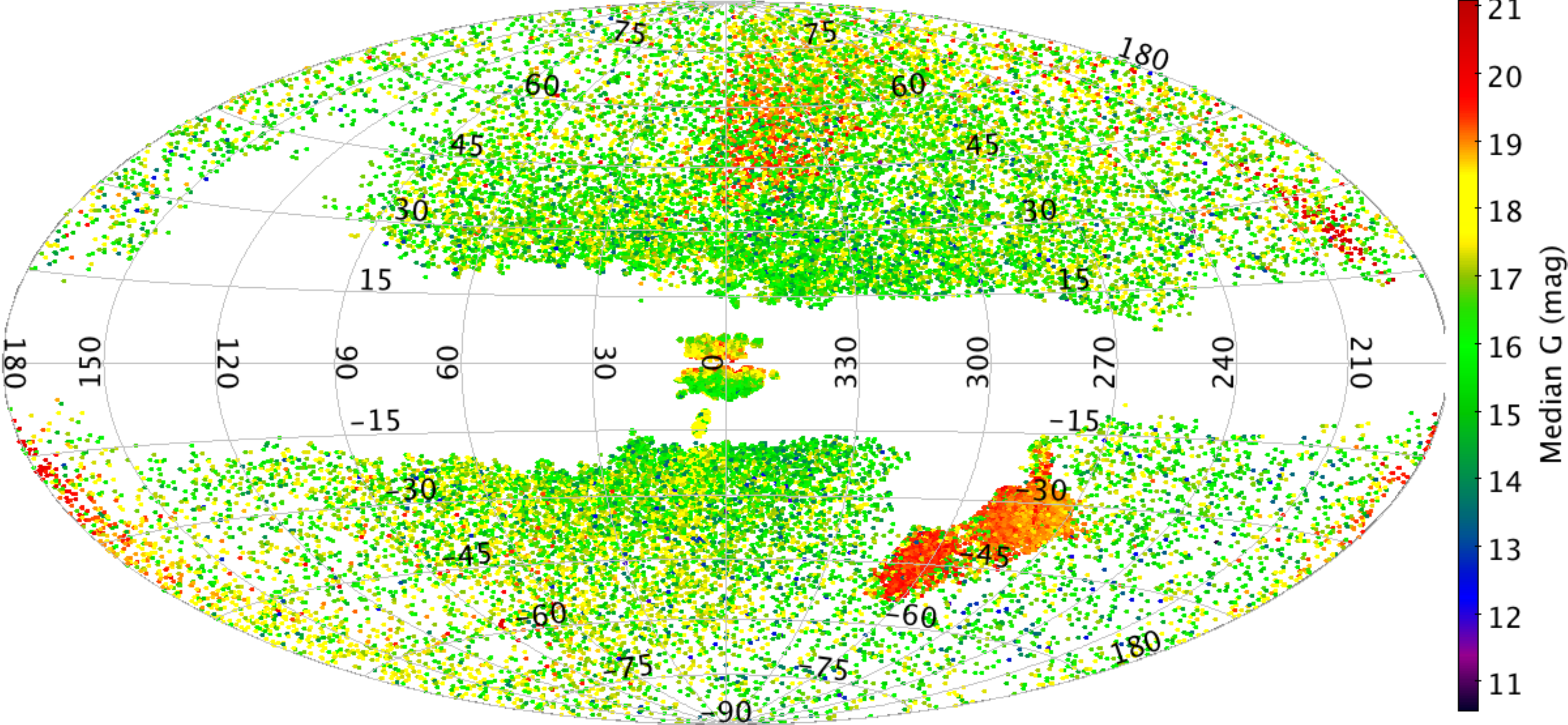}
\caption{Sky distribution in Galactic coordinates of Gaia cross-matched RR~Lyrae stars coloured in apparent magnitude (from the \href{https://www.cosmos.esa.int/web/gaia/iow_20170324}{Gaia Image of the Week of 24/03/2017}).}
\label{fig:RRABSkyPosition}       
\end{figure*}

In Figure~\ref{fig:RRABSkyPosition}, we present the sky distribution of the cross-matched RR~Lyrae stars, colour-coded with apparent magnitude.
We see clearly the Magellanic clouds, and the Sagittarius stream, and also the clump of distant and RR~Lyrae stars below the bulge at Galactic latitude $-14$~deg and longitude $+6$~deg.

Certainly a dedicated study should be done, in order to reach a precise description in the three spatial dimensions.
The addition of kinematical information would be even more interesting to decipher the origin of these structures.
The question of how far in time a halo configuration can be rolled back to understand in detail the formation process of the Milky Way is under debate.
\cite{HelmideZeeuw2000} showed the potential of astrometric missions to unravel the formation history of our Galaxy.
However \cite{JeanBaptisteEtAl2016} argued that the rolling back cannot be done too far in the past, because of the non-conservation of the integral of motions (probably due to  dynamical friction) and the pollution from the stars born in the disk which were diffused later by the accreted satellite.
Whatever the result of such theoretical studies, the detailed description of the Milky Way halo should be done with great detail.
It is very important to gather individual chemical abundances to add to the positional and kinematical descriptions.
The Gaia-ESO \cite{GilmoreEtAl2012}, Apogee \cite{AllendePrieto2008}, and future spectroscopic surveys are very important in this respect.

\subsection{The Gaia mission extension}
Initially the mission was planned for half a year of commissioning and 5~years of nominal science operations plus possibly one year of extension.
Obviously a longer mission means more data and more precise results.
For example the precision on parallax and proper motion for Gaia is proportional to $1/\sqrt{T}$, and $1/\sqrt{T^3}$ respectively, where $T$ is the length of the mission.
This improvement of the precision is obviously not applicable when a floor of systematic effects is reached, which will most likely be the case for the brightest ($\sim$1\%) subset of the targets.

A document was coordinated by the Gaia project scientist, Timo Prusti, to make the case of a longer mission, i.e., an extension up to 3-5~years.
While the mission lifetime limitation is due most likely to fuel for the micro-propulsion system, a 5-year extension could be achievable with the current fuel consumption and sunshield ageing conditions.
These coming years, the mission will go every second year through the ESA usual procedure to decide about the iterative 2-year extension of space missions.

	\section{Conclusions}\label{sect:conclusions}

In this article, we presented the Gaia mission and the content of the first data release.
We also explored some aspects of the future Gaia data releases in the perspective of pulsating star research.

The intent of this first release for variable stars was more to present a showcase of some capabilities of Gaia.
The peculiar ecliptic pole sampling of Gaia produces a number of measurements which is similar to the one reached in the nominal scanning law for 5 years. 
Even if the number of released light curves was small we had to go through all the mechanisms and procedures of an official release of the Gaia consortium and ESA.
This work was very instructive and demonstrates that the system (DPAC/ESA) is flexible enough to allow some results to be available publicly even though it was not initially planned.
Furthermore, as presented in the previous sections, there were already numerous articles published related to variable stars.

Obviously the first data release is just a very tiny tip of the iceberg, an iceberg that will reveal itself through the future releases. 
We can confidently conclude that Gaia will be an exceptional survey for pulsating star research.


\begin{acknowledgement} 
\noindent\vskip 0.2cm
\noindent {\em Acknowledgments}: 
  On February 28 2017, Jan Cuypers passed away unexpectedly.
  As co-authors we are filled with sadness, and we will miss our colleague dearly.
  At the same time we feel grateful for all the valuable contributions he made to this work, which will undoubtedly leave a lasting mark on all achievements of CU7 we can expect in the coming years.
  The authors would therefore like to dedicate this paper to Jan. 
  We thank Jos de Bruijne for valuable comments, Yves Revaz and Misha Haywood for discussions and Lennart Lindegren for a list of limitations of first Gaia data release.
  This work has made use of data from the ESA space mission {\it Gaia}, processed by the {\it Gaia} Data Processing and Analysis Consortium (DPAC). We acknowledge with thanks the variable star observations from the AAVSO International Database contributed by observers worldwide and used in this research.
\end{acknowledgement}

\bibliography{local}

\end{document}